\documentclass[aps,prl,twocolumn,showpacs,letterpaper,nofootinbib]{revtex4-1}
\usepackage{graphicx,amsmath,amstext}
\usepackage[outdir=./]{epstopdf}
\usepackage{color}
\usepackage{natbib}
\usepackage[normalem]{ulem}

\begin{document}

\title{Direct Measurement of Photon Recoil from a Levitated Nanoparticle}



\author{Vijay Jain$^{1,2}$, Jan Gieseler$^{1}$, Clemens Moritz$^{3}$, Christoph Dellago$^{3}$, Romain Quidant$^{4,5}$, and Lukas Novotny$^{1}$}

\homepage{www.photonics.ethz.ch}

\affiliation{$^1$ ETH Z\"{u}rich, Photonics Laboratory, 8093 Z\"{u}rich, Switzerland.}
\affiliation{$^2$ University of Rochester, Department of Physics and Astronomy, Rochester, NY 14627, USA.}
\affiliation{$^3$ University of Vienna, Faculty of Physics, Boltzmanngasse 5, 1090 Vienna, Austria.}
\affiliation{$^4$ ICFO-Institut de Ciencies Fotoniques, The Barcelona Institute of Science and Technology, 08860 Castelldefels (Barcelona), Spain.}
\affiliation{$^5$ ICREA-Instituci{\'o} Catalana de Recerca i Estudis Avan\c{c}ats, 08010 Barcelona, Spain.}

\date{\today}

\begin{abstract}
The momentum transfer between a photon and an object defines a fundamental limit for the precision with which the object can be measured.
If the object oscillates at a frequency $\Omega_0$, this measurement back-action adds quanta $\hbar\Omega_0$ to the oscillator's energy at a rate $\Gamma_{\rm recoil}$, a process called photon recoil heating, and sets bounds to coherence times in cavity optomechanical systems.
Here, we use an optically levitated nanoparticle in ultrahigh vacuum to directly measure $\Gamma_{\rm recoil}$. By means of a phase-sensitive feedback scheme, we cool the harmonic motion of the nanoparticle from ambient to micro-Kelvin temperatures and measure its reheating rate under the influence of the radiation field.  The recoil heating rate is measured for  different particle sizes and for different excitation powers, without the need for cavity optics or cryogenic environments. The measurements are in quantitative agreement with theoretical predictions and provide valuable guidance for the realization of quantum ground-state cooling protocols
and the measurement of ultrasmall forces.\\
\end{abstract}

\pacs{42.50.Wk, 62.25.Fg, 07.10.Pz}

\maketitle

%
Our ability to detect ultraweak forces depends on both the noise and sensitivity of the measurement. An optical position sensor, for example, irradiates an object with light and detects the scattered photons. As each photon carries momentum $p = \hbar k$, we can increase the optical power to reduce the object's position uncertainty to $\Delta{x} \ge 1/(2k\sqrt{N})$, where $N$ is the number of scattered photons. Increasing the optical power, however, increases the rate of momentum kicks from individual photons and results in a force due to radiation pressure shot noise (RPSN), which perturbs the inspected object. While increasing power reduces our measurement imprecision, RPSN places limits on the information gained from a system~\cite{braginsky78,caves81}.\\[-1.1ex]
	
	Remarkable advances in micro-fabrication have resulted in high-Q mechanical resonators required for enhanced force sensitivity. In addition to 
	ground state cooling~\cite{chan11a,verhagen12}, recent experiments in cryogenic chambers with silicon nitride membranes, cold-atomic clouds, and 
	microwave devices have verified the influence of RPSN in continuous position and force measurements~\cite{purdy13,schreppler14,wilson15,teufel15,peterson16}. Increasing the circulating optical power in the cavity increases the back-action to the resonator, which is manifested as an increase in the oscillator's mean-square displacement.\\[-1.1ex]

While cavity optomechanical systems seek to operate in this shot-noise dominant regime in order to observe macroscopic quantum phenomena, material impurities limit the quality factors within systems that are mechanically clamped to the environment. Furthermore, absorption of radiation limits the number of photons that can be used to interrogate the system.
Thus, despite cryogenic temperatures, thermal dissipation is often the dominant decoherence mechanism and places a material limit to the sensitivity of the device~\cite{wilson15}.\\[-1.1ex]
	
	Optically levitated nanoparticles in vacuum have proven to be versatile platforms for studies of light-matter interactions~\cite{ashkin70,gieseler12,yin13,kiesel13,millen15,ranjit15}. Free from mechanical vibrations of the environment, they have been used to investigate nonequilibrium fluctuation theorems~\cite{gieseler14}, nonlinear dynamics and synchronization~\cite{gieseler14b}, rotational motion~\cite{arita15}, ultrasmall forces~\cite{gieseler13a,ranjit15}, and coupling to internal spin degrees of freedom~\cite{neukirch15a}.
	 In the context of cavity optomechanics, levitated nanoparticles have also been proposed for quantum ground state cooling~\cite{chang10a,romeroisart11a,kiesel13,millen15} and for gravitational wave detection~\cite{kaltenbaek12,asima13}. Central to all of these experiments is the optical gradient force, which is needed to trap and control the scrutinized nanoparticle. However, due to the discrete nature of optical radiation, the trapping force is itself intrinsically noisy and RPSN may influence the motion of the trapped particle via photon recoil heating, akin to atomic physics where it limits the temperature of Sisyphus cooling to a few micro-Kelvins~\cite{tannoudji98}.\\[-1.1ex]
	
	In most situations photon recoil heating is negligibly small for macroscopic objects, as the recoil energy $\hbar^{2} k^{2}/(2m)$ scales inversely with the object's mass. Consequently, to observe this  weak effect, the system has to be sufficiently well isolated. In particular, the photon recoil rate has to be larger than the thermal decoherence rate.
	Using active feedback to bring a nanoparticle into ultrahigh vacuum (UHV) ($P_{\rm gas} \sim 10^{-8}$~mbar), however, we significantly reduce the heating due to residual gas molecules and thereby ascertain for the first time a direct readout of the recoiling rate of photons from a macroscopic object at room temperature. This places us in the regime of strong measurement back-action~\cite{teufel15}. \\[-1.1ex]

\begin{figure}[!b]
\begin{center}
\includegraphics[width=0.34\textwidth,angle=0]{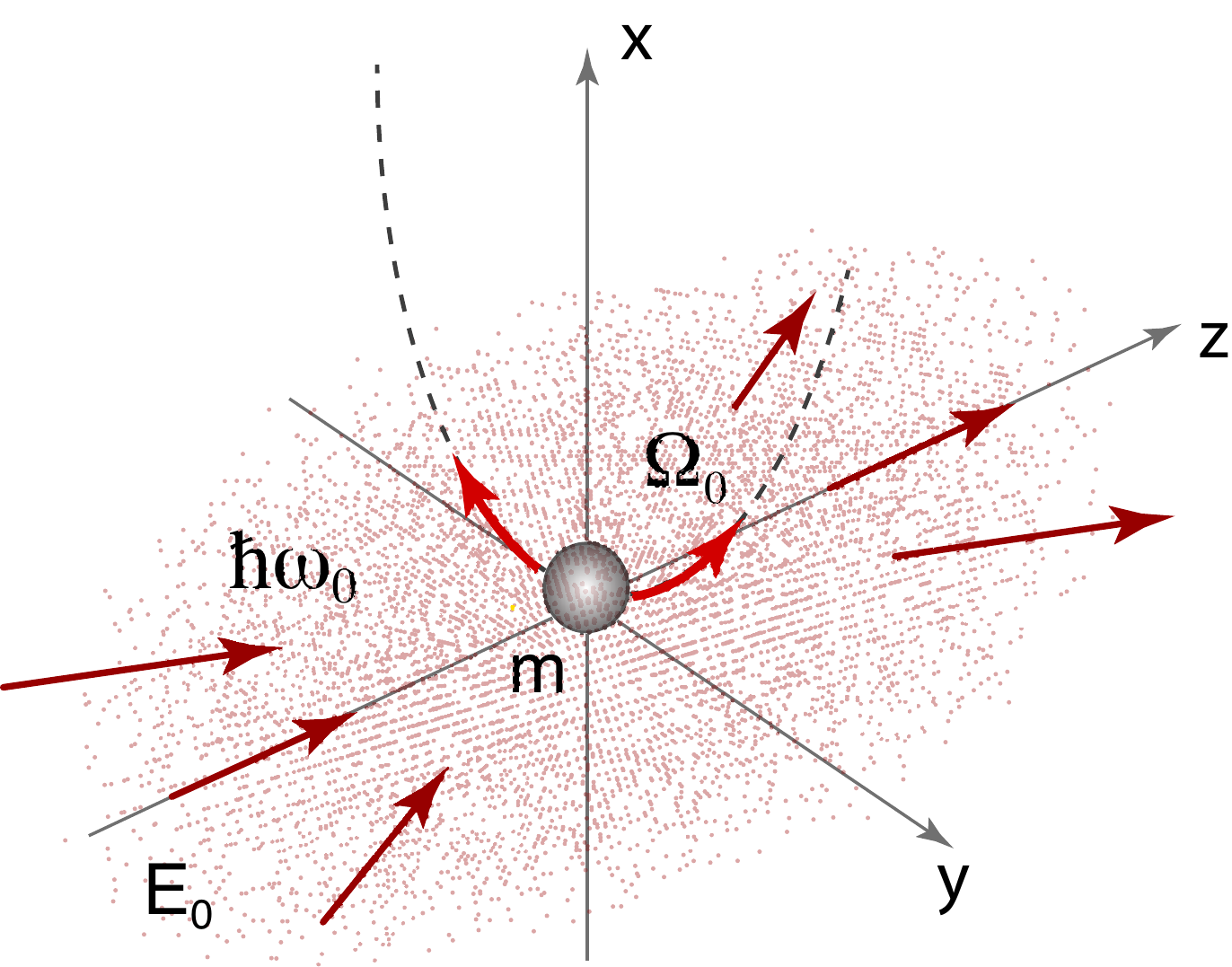}
\end{center}
\caption{{\bf Illustration of photon recoil heating.}  
A particle with mass $m$ is trapped at the focus of a laser beam by means of the optical gradient force. 
The particle's center-of-mass temperature is cooled by parametric feedback and heated by individual photon momentum kicks.
$\Omega_0\,/\,2\pi$ is the mechanical oscillation frequency and $\hbar \omega_0$ is the photon energy. The incident light is polarized along the $x$ direction. 
}
\label{fig01}
\end{figure}

To minimize the damping due to residual gas molecules we perform our experiments in UHV environments. As described in Ref.~\cite{gieseler12}, we first trap a particle by a strongly focused laser beam at ambient temperature and gas pressure, and then  evacuate the vacuum chamber. We use fused silica particles with radii on the order of $R = 50\,$nm, a laser beam with wavelength $\lambda=1064\,$nm and power $P_0=70\,$mW, and an objective of numerical aperture ${\rm NA}=0.9$ for focusing. We choose a coordinate system whose $z$ axis coincides with the optical axis and whose $x$ axis defines the direction of polarization of the incident light (c.f. Fig.~\ref{fig01}). \\[-1.1ex]

\begin{figure}[!t]
\begin{center}
\includegraphics[width=0.3\textwidth,angle=0]{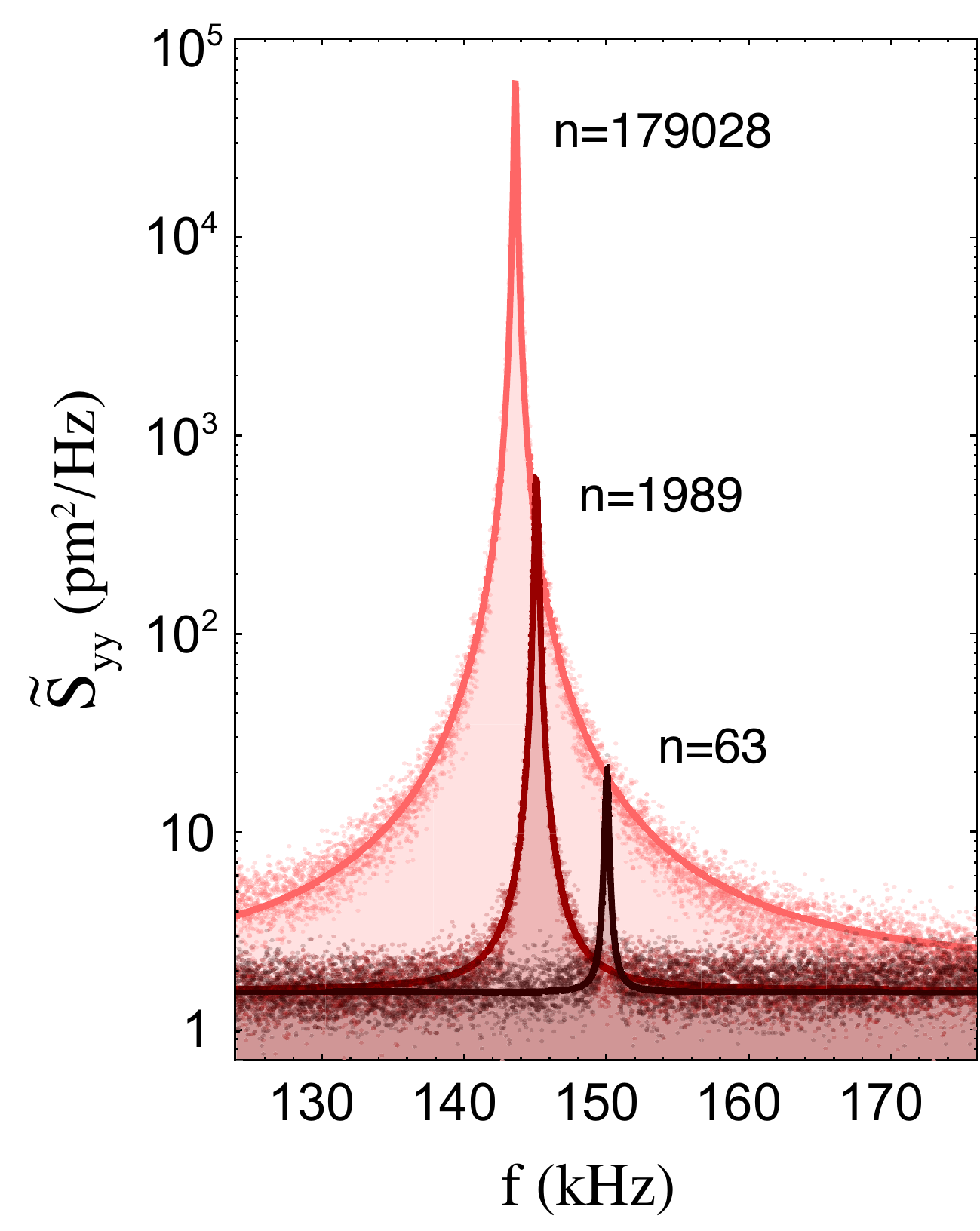}
\quad \hspace{1em}
\end{center}
\vspace{-1.5em}
\caption{{\bf Power spectral densities under feedback cooling}. The Lorentzian curves correspond to the motion of a particle with radius $R=49.8\,$nm along $y$ for three different vacuum pressures: $6.6\times10^{-4}$, $1.1\times10^{-5}$ and $2\times10^{-8}\,$mbar. ${\rm n}$ indicates the mean occupation number. The center-of-mass temperature of the  ${\rm n}=63$ peak is $T_{\rm cm}=450\,\mu$K. Note that $\tilde{S}_{yy}$ is the single sided PSD~\cite{psdnote}.
}
\label{fig02}
\end{figure}

For small oscillation amplitudes, the particle's motion along the three principal axes is decoupled and we end up with three independent harmonic oscillators, each with their own oscillation frequency $\Omega_0$ and damping $\gamma$, a result of the asymmetric shape of the optical potential~\cite{gieseler12}.  For example, the motion along $y$ is described by
\begin{equation}
\ddot y + \gamma\:\!\dot y + \Omega_0^2\:\! y\;=\; \frac{1}{m} F(t) \; ,
\label{motion}
\end{equation}
with $\Omega_0 / 2\pi = 150\,$kHz (c.f. Fig.~\ref{fig02})  and $F$ denoting fluctuating forces acting on the particle. The corresponding oscillation frequencies for the $x$ and $z$ axes are $123\,$kHz and $49\,$kHz, respectively. The oscillator's damping rate can be written as  $\gamma=\gamma_{\rm th} + \gamma_{\rm rad} + \gamma_{\rm fb}$, where $\gamma_{\rm th}$ accounts for the interaction with the background gas, $\gamma_{\rm rad}$ for the interaction with the radiation field, and $\gamma_{\rm fb}$ is the damping introduced by feedback cooling. The different contributions will be discussed in detail. \\[-1.1ex]

The trapped particle's energy changes constantly due to interactions with its environment and the time evolution of its average energy $\bar E$ is predicted by the Fokker-Planck equation to be~\cite{gieseler14}
\begin{eqnarray}
\frac{d}{dt} \bar E(t) \;=\; -\gamma \left[ \bar E (t)\,-\,E_{\infty}\right]  \; ,
\label{theory01}
\end{eqnarray}
where $E_{\infty}$ is the average energy in the steady state ($t\to \infty$) and $\gamma$ is the rate at which the steady state is being reached. Writing the average energy of the particle in terms of discrete quanta, $\bar E = {\rm n} \, \hbar\Omega_0$, we  obtain
\begin{eqnarray}
{\rm \dot n} \;=\; -\gamma \;\!  {\rm n}\,+\, \Gamma\; ,
\label{theory02}
\end{eqnarray}
where ${\rm n}$ is the mean occupation number and
\begin{eqnarray}
\Gamma \,=\,\!\frac{E_{\infty}}{\hbar\Omega_0}\,\gamma  \; .
\label{theory03}
\end{eqnarray}
is the heating rate. It defines the rate at which phonons are reintroduced into the mechanical system.
The solution of Eq.~(\ref{theory02}) is
\begin{eqnarray}
{\rm n}(t) \;=\; {\rm n_{\infty}}\, + \left[{\rm n_0}-{\rm n_{\infty}}\right]\,{\rm e}^{-\gamma  t}\;,
\label{theory04}
\end{eqnarray}
 where ${\rm n_0}$  is the mean occupation number at an initial time and
\begin{eqnarray}
{\rm n_{\infty}} \;=\; \frac{\Gamma}{\gamma}\;
\;=\; \frac{\Gamma_{\rm th}+\Gamma_{\rm recoil}+ \Gamma_{\rm fb}}{\gamma_{\rm th} + \gamma_{\rm rad} + \gamma_{\rm fb}}\;
\label{theory04a}
\end{eqnarray}
 is the occupation number in the steady state.
In \eqref{theory04a} we have written $\Gamma$ as  the sum of a heating rate due to collisions with gas molecules ($\Gamma_{\rm th}$), a heating rate due to photon recoil kicks ($\Gamma_{\rm recoil}$), and a heating rate due to noise introduced by the feedback loop ($\Gamma_{\rm fb}$).\\[-1.1ex] 

The surrounding gas at temperature $T$ gives rise to damping $\gamma_{\rm th}$ and  thermal decoherence $\Gamma_{\rm th} = \gamma_{\rm th}\, k_B T\left/\hbar \Omega_0\right.$. For $\gamma > \Omega_0$, the particle's motion is overdamped and the dynamics are governed by a diffusion equation, as in the case of optical tweezers operated in liquids.
At gas pressures below $10\;{\rm mbar}$, the damping to the nanoparticle is linear in gas pressure~\cite{beresnev90}
\begin{equation}
\gamma_{\rm th} \,\approx\,  15.8\,\frac{R^{2} P_{\rm gas}}{m \:\!v_{\rm gas}} \, ,
\label{gas}
\end{equation}
where $P_{\rm gas}$ is the pressure, $v_{\rm gas}=\sqrt{3 k_B T / m_{\rm gas}}$ and $m_{\rm gas}$ are the root-mean-square velocity and mass of gas molecules, and $R$ and $m$ are the particle's radius and mass, respectively.\\[-1.1ex]

%
Left alone, the trapped particle will have $n_{\rm th} = k_B T\left/ \hbar \Omega_0\right.$ thermal quanta on average.
However, by means of parametric feedback \cite{gieseler12} we introduce a cold damping $\gamma_{\rm fb}$, which 
cools the particle to occupation numbers much lower than $n_{\rm th}$.
The feedback consists of a split detection scheme in combination with a phase-locked loop (PLL) for phase sensitive detection of the particle's motion and feedback control. 
As shown in Fig.~\ref{fig02}, by means of feedback cooling we are able to reach mean occupation numbers of $n=62.5 \pm 5$, which corresponds to a center-of-mass temperature of  $T_{\rm cm}=(450.5 \pm 33.1)\!\:\mu$K. \\[-1.1ex]

%
At very low pressures, $\Gamma_{\rm th}$ becomes negligibly small and,  in absence of feedback cooling, the particle's heating is dominated by photon shot noise, {\it i.e.} the random momentum kicks imparted by photon scattering.
Photon recoils imparted to the nanoparticle give rise to  radiation pressure back-action, that is, a disturbance of the particle's motion. The power spectral density (PSD) of the displacement along the $y$ direction is
\begin{eqnarray}
S_{yy}(\Omega) \;=\;  \left|\chi(\Omega)\right|^2  \,S_{yy}^F \; ,
\label{theory10}
\end{eqnarray}
where $\chi(\Omega)= (1/m)/(\Omega_0^2-\Omega^2 - i\gamma \Omega)$ is the susceptibility (transfer function) of the harmonic oscillator and $S_{yy}^F$ is the power spectral density of the force acting on the nanoparticle.  In the  limit of a negligible contribution from the residual gas, $S_{yy}^F$ is dominated by photon shot noise, {\it i.e.}~\cite{suppl}
\begin{eqnarray}
S_{yy}^F  \;=\;  \frac{2}{5}  \frac{\hbar\omega_0}{2\pi c^2} \, {P}_{{\rm scatt}} \; .
\label{theory11}
\end{eqnarray}
Here, $\hbar\omega_0$ is the photon energy and $P_{\rm scatt}$ is the scattered power of the particle. The mean-square displacement is calculated as
\begin{eqnarray}
\langle y^2 \rangle  \;=\;  \int_{-\infty}^{\infty} \!S_{yy}(\Omega)\; d\Omega\; \;=\; \frac{1}{5} \frac{\hbar\omega_0}{m\:\! \Omega_0^2} \frac{P_{\rm scatt}}{\,m\:\! c^2} \frac{1}{\gamma}\; .\;\;\;
\label{theory12}
\end{eqnarray}
Assuming that the particle attains a thermal steady state, we invoke the equipartition theorem $\hbar \Omega_0 n_{\infty} = K_s\:\! \langle y^2\rangle$, with trap stiffness $K_s = m \Omega_0^2$.
Inserting this expression into (\ref{theory03}) we finally find the recoil heating rate to be 
\begin{equation}
\Gamma_{\rm recoil}\;=\; \frac{1}{5}\;\!\frac{P_{\rm scatt}}{\,m\:\!c^2}\:\! \frac{\omega_0}{\Omega_0}\; ,
\label{theory14}
  \end{equation}
in agreement with atomic theory~\cite{itano82,chang10a}. Note that similar results are obtained for the displacements in $x$ and $z$ directions, but with different oscillation frequencies $\Omega_0$. For the $z$ direction the recoil formula turns out to be identical to (\ref{theory14})  whereas for the $x$ direction (along the polarization axis) it is only half as large.\\[-1.1ex]

\begin{figure}[!t]
\begin{center}
\hspace{-2em} \includegraphics[width=0.36\textwidth]{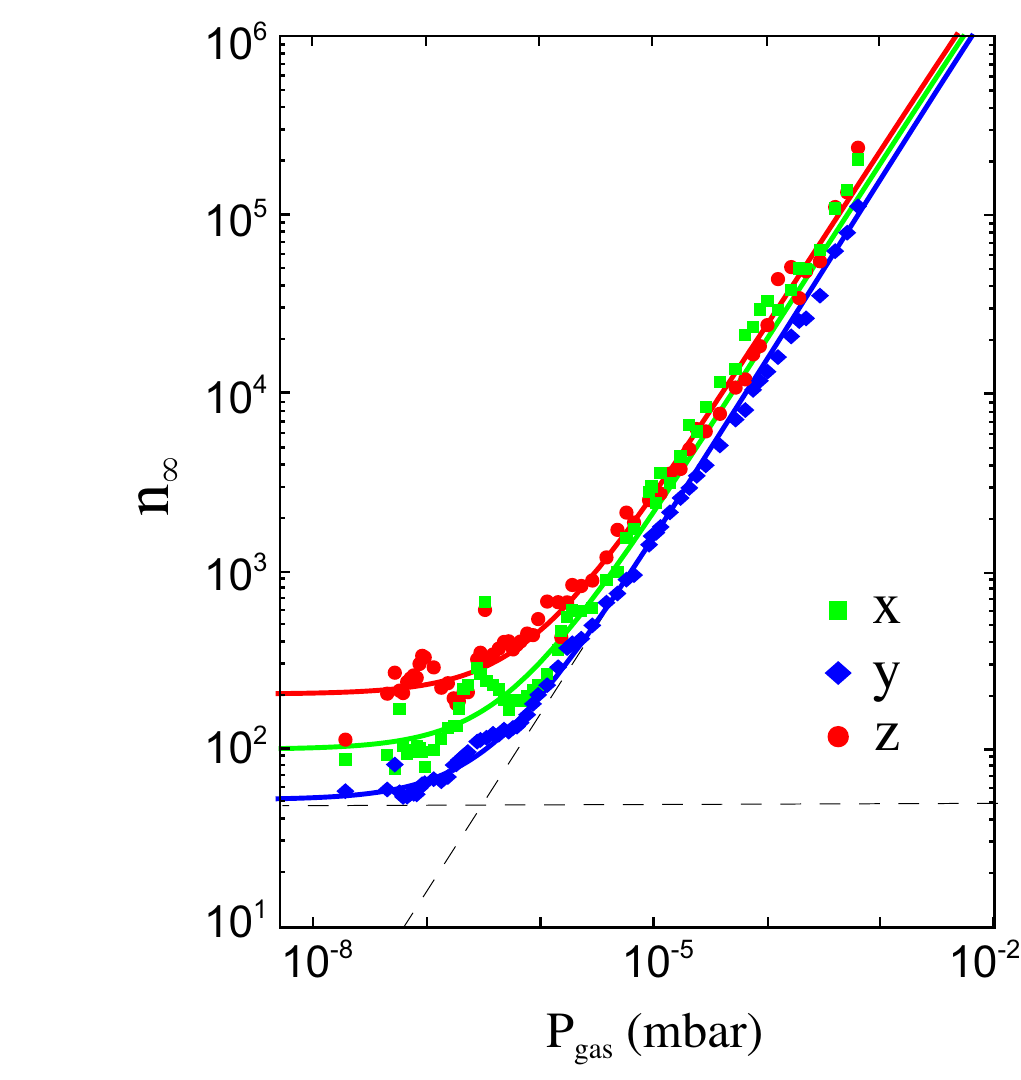}
\end{center}
\vspace{-1.5em}
\caption{{\bf Steady-state  under feedback cooling.} Mean occupation number along the three principal axes (x, y, z) as a function of gas pressure measured under constant feedback cooling for a $R = 49.8~{\rm nm}$ particle with focal power $P_{0} = 70~{\rm mW}$.
	 The solid curves are fitting functions of the form $a + b P_{\rm gas}$. 
}
\label{fig03}
\end{figure}

Let us estimate the magnitude of $\Gamma_{\rm recoil}$. For a Gaussian beam, the intensity at the laser focus is $I_0=P_0  k^2 {\rm N\!A^2} \:\!/\:\! 2 \pi$, where $k=\omega_0/c$. The scattering cross-section is derived from the particle polarizability $\alpha$ as $\sigma_{\rm scatt} = |\alpha|^2 k^4 \:\!/ \:\!6\pi \varepsilon_0^2$, where $\alpha = 4\pi\varepsilon_0 R^3 (n^2-1)/(n^2+2)$, $n$ is the index of refraction and $R$ the particle's radius. The scattered power is then calculated as $P_{\rm scatt}=\sigma_{\rm scatt} I_0$. For the parameters used in Figs.~\ref{fig02} and~\ref{fig03} ($n=1.45$, $\lambda=1064\,$nm, $P_0=70\,$mW, $R = 49.8\,$nm, ${\rm N\!A}=0.9$) we find $P_{\rm scatt}=3.53\,\mu$W. The specific mass density of silica is $\rho_{\rm SiO_2}=2200\,$kg/m$^3$ and the mass of the particle turns out to be $m= 1.14\times 10^{-18}\,$kg. Using $\Omega_0= 2\pi \times150\,$kHz, Eq.~(\ref{theory14}) predicts a reheating rate of $\Gamma_{\rm recoil}= 13.0\,$kHz. \\[-1.1ex]

In addition to heating, the radiation field also leads to radiation damping at a rate $\gamma_{\rm rad}$, which arises from the Doppler effect~\cite{karrai08a} and can be evaluated by calculating the back-action of the scattered field on the motion of the particle along the $y$ axis. We find a value of  $\gamma_{\rm rad} \sim P_{\rm scatt} / m c^2$.  
Note that in the photon dominated regime and in the absence of feedback cooling, the equilibrium temperature  $k_B T_{\infty} = \hbar \Omega_0 n_{\infty}\sim \hbar \omega_0$ is of the order of the photon energy. This energy is comparable to the depth of the trapping potential in our experiments and therefore the particle is likely to escape as it heats up without feedback control. \\[-1.1ex]

%
To experimentally verify the departure from the gas dominated heating regime, we record the particle's average energy $\bar E$ as a function of gas pressure $P_{\rm gas}$ under constant feedback cooling. The result is shown in Fig.~\ref{fig03} where we expressed the average energy in terms of the mean occupation number $n_{\infty}= \bar E / \hbar\Omega_0$.
 The figure demonstrates that as pressure is reduced to $10^{-7}$~mbar, the gas damping attenuates linearly with pressure, in agreement with Eq.~(\ref{gas}).
At pressures lower than $10^{-7}\,$mbar, however, 
the particle's motion is weakly influenced by interactions with the gas. In this regime the dynamics are primarily determined by particle-photon interactions and the feedback loop. \\[-1.1ex]

The center-of-mass temperature scales with the integral of the power spectral density (c.f.~Fig.~\ref{fig02}), while the width of the peak yields the damping $\gamma \simeq \gamma_{\rm fb}$.
For example, the Lorentzian peak labeled with ${\rm n}=63$  in Fig.~\ref{fig02}
corresponds to a center-of-mass temperature of  $T_{\rm cm}=(450.5 \pm 33.1)\!\:\mu$K and yields a damping of $\gamma_{\rm fb} = 2\pi\times 269.9 \rm Hz$. \\[-1.1ex]

We perform a direct measurement of the recoil rate in a ring up style measurement, whereby the feedback is switched off at $t=0$ and the particle is allowed to heat up. By inactivating the feedback we eliminate the contribution of feedback induced heating ($\Gamma_{\rm fb}$)~\cite{suppl}.
As described in Ref.~\cite{gieseler14}, individual reheating trajectories represent a stochastic process and, thus, the heating rate and temperature have to be extracted from averages over  many individual reheating trajectories.
After switching-off the feedback we follow individual reheating trajectories over time periods that are considerably shorter than $1/\gamma$, which allows us to linearize the exponential term in Eq.~(\ref{theory04}). We then obtain
\begin{eqnarray}
{\rm n}(t) \,=\,  {\rm n_0} - \gamma \left[{\rm n_0}- {\rm n_\infty}\right] \,t + ..\; \approx\, {\rm n_0} + \Gamma_{\rm recoil}\,t \, .\qquad 
\label{theory04b}
\end{eqnarray}
In the last step, we used the fact that ${\rm n_0}\ll {\rm n_\infty}$, a condition that is fulfilled in our experiments owing to feedback cooling. Thus, we find that the reheating is linear in time shortly after switching off the feedback and that the main contribution to the reheating rate is the photon recoil rate  $\Gamma_{\rm recoil}$. We extract $\Gamma_{\rm recoil}$ from our measurements and study it as a function of  system parameters, such as laser power, particle size, and gas pressure. \\[-1.1ex]

\begin{figure}[!b]
\begin{center}
	\hspace{-1.5em} \includegraphics[width=0.48\textwidth]{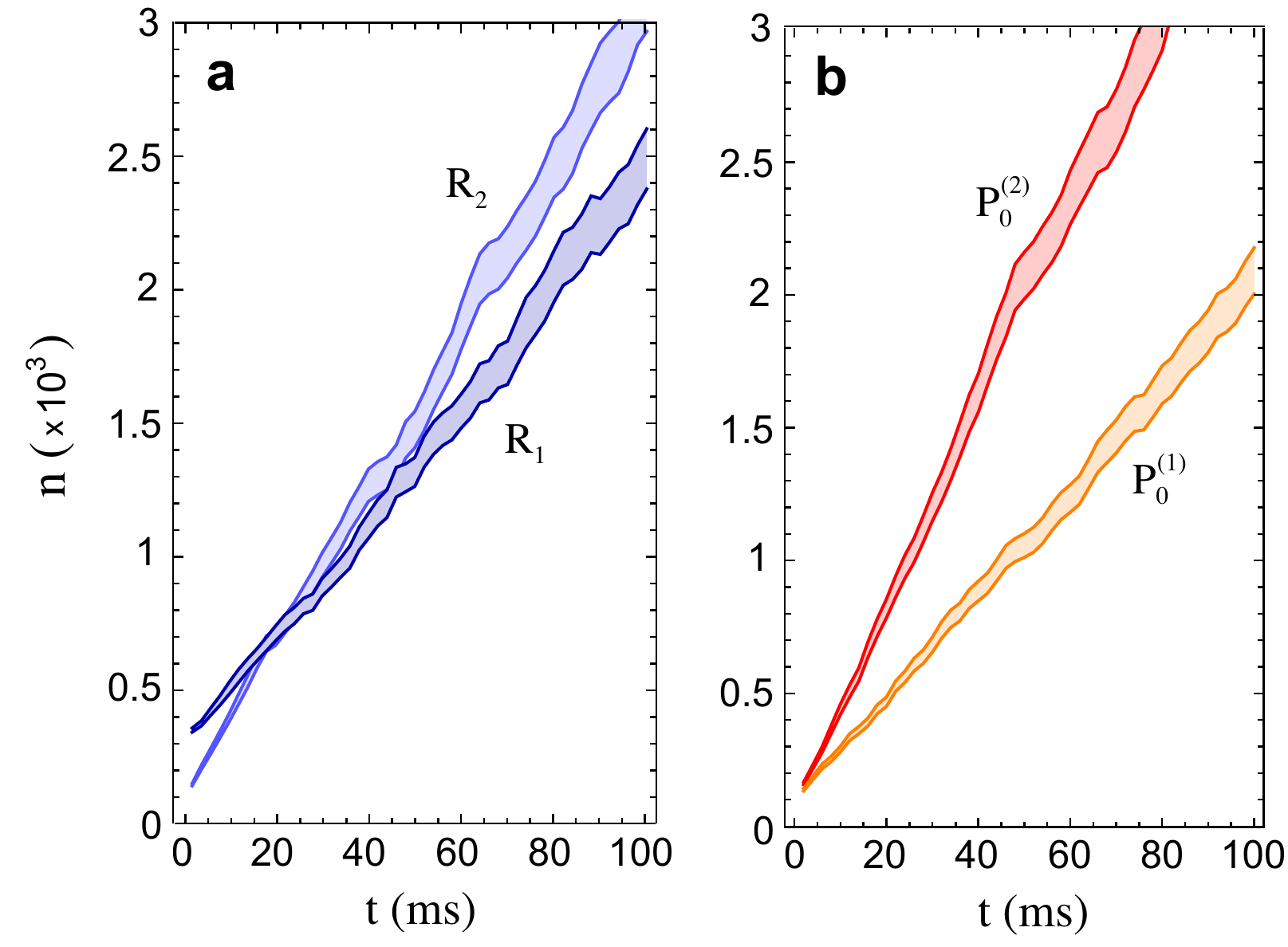}
\end{center}
\vspace{-1.5em}
\caption{{\bf Reheating time-traces.} Particle reheating along the $y$ axis for different particle sizes and laser powers. 
(a) Reheating for $R_{1} = 52.7\,{\rm nm}$ and $R_{2} = 71.6\,{\rm nm}$ nanoparticle. The pressure is $3\times 10^{-8} {\rm mbar}$ and the focal power is $70\,{\rm mW}$. (b) Reheating for a particle with radius $R = 68.0\, {\rm nm}$ measured for two different focal powers, $P_{0}^{(1)} = 30.5\, {\rm mW}$ and $P_{0}^{(2)} = 80.0\, {\rm mW}$, at a pressure $7 \times 10^{-9}\,{\rm mbar}$. The experimental data are obtained by averaging 500 individual reheating trajectories. The shaded areas reflect one standard error above and below the mean phonon value.  }
\label{fig04}
\end{figure}

Figure~\ref{fig04}a shows experimentally measured reheating time-traces for  two different particles with radii $R_{1} = 52.7\,{\rm nm}$ and $R_{2} = 71.6\,{\rm nm}$. The initial occupation number $n_{0}$ for the two particles is slightly different and the oscillation frequencies are $\Omega_{0}^{(1)} = 2\pi \times 148.8\,{\rm kHz}$ and $\Omega_{0}^{(2)} = 2\pi \times 151.2\,{\rm kHz}$. The slope of the time-traces  directly renders the reheating rate.  We obtain $\Gamma_1 = (20.9 \pm 0.2)\,{\rm kHz}$ and $\Gamma_2= (29.4 \pm 0.3)\,{\rm kHz}$. By comparison, the theoretical photon recoil rates according to~(\ref{theory14}) are $15.5\,{\rm kHz}$ and $38.2\,{\rm kHz}$, respectively.\\[-1.1ex]

We also measured the reheating rate as a function of focal power $P_{0}$. Fig.~\ref{fig04}b shows the reheating time-traces of a $R = 68.0\,{\rm nm}$ particle measured with  inferred laser powers $P_{0}^{(1)}= 30.5\,{\rm mW}$ and $P_{0}^{(2)} = 80\,{\rm mW}$. The oscillation frequencies for the two laser powers are $\Omega_{0}^{(1)} = 2\pi \times 100.5\,{\rm kHz}$ and $\Omega_{0}^{(2)} = 2\pi \times 158.8\,{\rm kHz}$ with corresponding reheating rates  $\Gamma_1= (19.4 \pm 0.1)\,{\rm kHz}$ and $\Gamma_2 = (38.0 \pm 0.3)\,{\rm kHz}$. By comparison,  the theoretical photon recoil rates are $21.5\,{\rm kHz}$ and $35.8\,{\rm kHz}$, respectively. Discrepancies between measured rates and theoretical predictions result from residual gas heating and our use of the paraxial approximation for the focused laser field.\\[-1.1ex]

Our measurements indicate that heating due to the shot noise of photons is the dominant dissipation mechanism in our system when the feedback is inactivated. As shown in Fig.~\ref{fig04}a, increasing the particle's size heats it up faster despite starting off with lower $\rm n$, and, as shown in Fig.~\ref{fig04}b, reducing laser power reduces the heating rate, both consistent with photon recoil heating described by Eq.~(\ref{theory14}).
%
%
In our experiments, decoherence due to photon shot noise overwhelms thermal decoherence by at least a factor of 25 in our experiments, a ratio that could be raised by further reducing our vacuum pressure. 
To the best of our knowledge, our experiments are the first direct measurement of the photon recoil rate from a mesoscopic object at ambient temperatures. We find that for nanoscale particles $\Gamma_{\rm recoil}$ is approximately $10\,$kHz, which sets limits to ground-state cooling protocols,  and limits the maximum achievable quality factors and force sensitivities. \\[-1.1ex]

\begin{acknowledgments}
This research was supported by ERC-QMES (no. 338763). VJ was supported in part by an NSF Graduate Research Fellowship. (no. DGE-1419118). 
RQ was supported by ERC-QnanoMECA (no. 64790), FIS2013-46141-P and Fundaci{\'o} Privada CELLEX. CD was supported by the Austrian Science Fund (FWF) within the SFB ViCoM (grant F41). CM was supported by a uni:docs-fellowship of the University of Vienna.  The authors thank  M.~Frimmer, E.~Hebestreit, R.~Reimann, and L.~Rondin for stimulating discussions. 
\end{acknowledgments}


%

\newpage

\def\tensor#1{\stackrel{\:\!\leftrightarrow}{\bf #1}}

\begin{widetext}

\quad \vspace{0em}

\begin{center}
\noindent {\large Supplementary Information}\\[4ex]

\noindent {\bf \Large Direct Measurement of Photon Recoil from a Levitated Nanoparticle}\\[2.5em]

{\rm Vijay Jain$^{1,2}$, Jan Gieseler$^{1}$, Clemens Moritz$^{3}$, Christoph Dellago$^{3}$, Romain Quidant$^{4,5}$ and Lukas Novotny$^{1}$}\\[3ex]

{\small \it
\begin{enumerate}
\item{ETH Z\"{u}rich, Photonics Laboratory, 8093 Z\"{u}rich, Switzerland.}\\[-4ex]
\item{University of Rochester, Department of Physics and Astronomy, Rochester, NY 14627, USA.}\\[-4ex]
\item{University of Vienna, Faculty of Physics, Boltzmanngasse 5, 1090 Vienna, Austria.}\\[-4ex]
\item{ICFO-Institut de Ciencies Fotoniques, The Barcelona Institute of Science and Technology, 08860 Castelldefels (Barcelona), Spain.}\\[-4ex]
\item{ICREA-Instituci{\'o} Catalana de Recerca i Estudis Avan\c{c}ats, 08010 Barcelona, Spain.}\\[6em]
\end{enumerate}
}
\end{center}

\section{1. Particle displacement due to photon scattering}
The force fluctuations acting on a nanoparticle can be expressed in terms of a correlation function
\begin{eqnarray}
\left\langle F_i(t)\,F_j(t+t')\right\rangle  = \lim_{T\rightarrow\infty} \frac{1}{T} \int_{-T/2}^{T/2} \!\!\!F_i(t)\,F_j(t+t')\, dt \, ,\;\;\;\;\;
   \label{coh01}
\end{eqnarray}
with $i,j\in\{x,y,z\}$. The spectral density  of these fluctuations follows from the Wiener-Khinchin theorem
\begin{eqnarray}
S_{F_{i}F_{j}}(\omega)&=& \int_{-\infty}^{\infty} \left\langle \hat F_i(\omega)\,\hat F^{\ast}_j(\omega')\right\rangle d\omega' 
\;=\; \frac{1}{2\pi} \int_{-\infty}^{\infty}  \left\langle F_i(t)\,F_j(t+t')\right\rangle {\rm e}^{i\omega t'} dt' \; ,\;\;\;
   \label{coh02}
\end{eqnarray}
where $\hat F_i(\omega)$ is the Fourier transform of $F_i(t)$.\\

The force acting on the particle is $F_i(\omega)=P_{\rm scatt}^{(i)}(\omega) / c$, with $P_{\rm scatt}^{(i)}$ being the power scattered in the direction $i$. Therefore, 
\begin{eqnarray}
\left\langle \hat F_i(\omega)\,\hat F^{\ast}_j(\omega')\right\rangle \;=\; \frac{1}{c^2}\,\left\langle \hat P_{\rm scatt}^{(i)}(\omega)\,\hat P^{\ast\,(j)}_{\rm scatt}(\omega')\right\rangle  .\;\;\;\;\;
\end{eqnarray}
If shot noise is the dominant source of fluctuations then the power spectral density  is~\cite{schottky18}
\begin{eqnarray}
S_{P_i P_j}(\omega)&=& \int_{-\infty}^{\infty} \left\langle \hat P_{\rm scatt}^{(i)}(\omega)\;\hat P^{\ast\,(j)}_{\rm scatt}(\omega')\right\rangle d\omega' 
\;=\; \frac{\hbar\omega_0}{2\pi}\, P_{\rm scatt}^{(i,j)}(\omega) \; ,
   \label{coh03a}
\end{eqnarray}
with $\hbar\omega_0$ being the photon energy and 
\begin{eqnarray}
P_{\rm scatt}^{(i,j)}(\omega)= P_{\rm scatt} \int_{0}^{2\pi}\!\!\!\!\!\int_{0}^{\pi}\!\! f(\theta,\phi)\, i(\theta,\phi)\,  j(\theta,\phi)  \,\sin\theta \,d\theta\,d\phi .\nonumber \\[-1ex]
   \label{coh04}
\end{eqnarray}
Here, $P_{\rm scatt}$ is the total scattered power, $f(\theta,\phi) = (3\,/\,8\pi) \sin^2\!\theta$ is the radiation pattern of an $x$ oriented dipole, and $i,j$ are components of the unit vector  ${\bf n} = [\cos\theta,  \sin\theta \cos\phi, \sin\theta \sin\phi]$. Evaluating the integral yields
$P_{\rm scatt}^{(x,x)}=(1/5) P_{\rm scatt}$, $P_{\rm scatt}^{(y,y)}=(2/5) P_{\rm scatt}$ and  $P_{\rm scatt}^{(z,z)}=(2/5) P_{\rm scatt}$.  The components with $i\neq j$ are zero. The spectral density along the $y$ direction now becomes 
\begin{eqnarray}
S_{yy}^F(\omega)\;=\; \frac{2}{5} \frac{\hbar\omega_0}{2\pi c^2} \, P_{\rm scatt}(\omega) \; ,  \label{coh05}
\end{eqnarray}
with similar expressions for $S_{xx}^F$ and $S_{zz}^F$. \\[2ex]

\section{2. Heating due to classical noise}
Classical noise associated with laser intensity fluctuations leads to a modulation of the trapping potential. It modifies equation~(1) in the main text into
\begin{eqnarray}
\ddot y + \gamma\:\!\dot y + \Omega_0^2\:\! y [1+ \epsilon(t)] \;=\; \frac{1}{m} F(t) \; ,
\end{eqnarray}
with the normalized white noise term $\epsilon(t)$, which gives rise to a heating rate of~\cite{bourret73,gehm98} 
\begin{eqnarray}
\Gamma_{\epsilon}\;=\; \pi\, \Omega_0^2 \,S^{\epsilon}\, {\rm n}
\;=\;\gamma_{\epsilon}\, {\rm n} \; .
\label{heatgamm}
\end{eqnarray}
Here, $S^{\epsilon}$ is the power spectral density of the noise, $\epsilon(t)$. \\

The heating rate~(\ref{heatgamm}) due to classical intensity fluctuations modifies the rate equation~(3) in the main text into
\begin{eqnarray}
{\rm \dot n} \;=\; -\gamma_{\rm rad} \;\!  {\rm n}\,+\, \Gamma_{\rm recoil}\,+\, \gamma_{\epsilon}\, {\rm n} \; ,
\label{theory02x}
\end{eqnarray}
where we assumed no gas heating. The solution is ${\rm n}(t) = {\rm n_{\infty}} + [{\rm n_0} - {\rm n_{\infty}}]\exp[-(\gamma_{\rm rad}-\gamma_\epsilon) t]$ with ${\rm n_{\infty}} = \Gamma_{\rm recoil}/(\gamma_{\rm rad}-\gamma_\epsilon)$. The linearized solution becomes 
\begin{eqnarray}
 {\rm n}(t) &=&  {\rm n_0} + \left[\Gamma_{\rm recoil}+\gamma_{\epsilon}n_0\right] t \; ,
\label{theory20x}
\end{eqnarray}
where we made use of $\gamma_{\epsilon}\gg \gamma_{\rm rad}$. Thus, the more the particle is cooled and the lower ${\rm n_0} $ is, the less it suffers from parametric heating due to laser intensity noise. Therefore, to overcome the limitations imposed by laser intensity fluctuations we require that the reheating experiments start out from a state of low occupation number. For example, using ${\rm n_0} =100$, together with the relative intensity noise (RIN) of the trapping laser of $-140\,$dB/Hz, we obtain $\gamma_{\epsilon} n_0 = 0.22\,$Hz, which is negligible compared to the recoil heating rate $\Gamma_{\rm recoil}\sim10\,$kHz.\\[2ex]

\section{3. Occupation number}
The mean thermal occupation number is
\begin{eqnarray}
n  \;=\;  \frac{k_B T}{\hbar\Omega_0} \;=\; \frac{m\:\! \Omega_0^2 \,\langle y^2\rangle}{\hbar \:\!\Omega_0} \, ,
\label{sqlfit04}
\end{eqnarray}
where we made use of the equipartition principle and where $\langle y^2\rangle$ is the particle's mean-square displacement. From the power spectral density we derive
\begin{eqnarray}
\langle y^2\rangle  \;=\;  \int_{-\infty}^{\infty} \!\!  {S}_y(\Omega)\; d\Omega \;=\;  \pi \gamma\,  {S}_y^{^{\rm peak}}\!(\Omega_0) \; .
\label{sqlfit02b}
\end{eqnarray}
Here, ${S}_y^{^{\rm peak}}\!$ denotes the amplitude of the Lorentzian peak measured against the imprecision background level. Combining the two equations yields
\begin{eqnarray}
n  \;=\;   \frac{\pi\:\!m\:\! \Omega_0\:\! \gamma}{\hbar}  \,{S}_y^{^{\rm peak}}\!(\Omega_0)  \; .
\label{sqlfit05}
\end{eqnarray}

\vspace{2ex}

For the parameters used in our experiments ($n=1.45$, $\lambda=1064\,$nm, $P_0=70\,$mW, $R=49.8\,$nm, ${\rm N\!A}=0.9$) we find $P_{\rm scatt}=3.53\,\mu$W. The mass of the particle amounts to $m= 1.14\times 10^{-18}\,$kg. Using $\Omega_0= 2\pi \times150.030\,$kHz, $\gamma = 2\pi\times 269.9\,$Hz, and  $\tilde{S}_y^{^{\rm peak}}\!=14.53\;{\rm pm^2 / Hz}$ \ ($S_y^{^{\rm peak}}\!=1.16\;{\rm pm^2 / Hz}$) \  we derive $n=63.0$. \\[-1.1ex]

The value of $\tilde{S}_y^{^{\rm peak}}\!=14.53\;{\rm pm^2 / Hz}$ corresponds to the single-sided PSD (see Section~\ref{sql}) and is the amplitude of the Lorentzian curve labeled with $n=63$ in Fig. 2 of the main text.\\

\section{4. Standard quantum limit\label{sql}}
On resonance $(\Omega=\Omega_0)$ and under feedback cooling the total power spectral density of the displacement noise  is
\begin{eqnarray}
S_y(\Omega_0)  \,=\,  S_y^{^{\rm imp}}\!(\Omega_0) +S_y^{^{\rm back}}\!(\Omega_0)+S_y^{^{\rm fb}}\!(\Omega_0)\, =\, 
\frac{S_y^{^{\rm zp}}\!(\Omega_0)}{2} \left[
\frac{1}{\eta_c}\,\frac{m c^2 \gamma\;\!\Omega_0}{2\:\!\omega_0\:\!P_{\rm scatt}}  \,+\,
\frac{2}{5}  \frac{2\:\!\omega_0\:\!P_{\rm scatt}}{m c^2 \gamma\;\!\Omega_0}\right] +\,S_y^{^{\rm fb}}\!(\Omega_0)  ,
 \label{sql14}
\end{eqnarray}
where $S_y^{^{\rm imp}}$ and $S_y^{^{\rm back}}$ are the power spectral densities of imprecision and backaction, respectively, $S_y^{^{\rm fb}}$ is the noise introduced by the feedback and $S_y^{^{\rm zp}}$ is the zero-point spectral density
\begin{eqnarray}
S_y^{^{\rm zp}}\!(\Omega_0) = \frac{\hbar}{2\pi\:\!m\:\!\gamma\:\!\Omega_0} \; ,
 \label{sql15}
\end{eqnarray}
with $\gamma$ being the damping. The parameter $\eta_c$ in Eq.~(\ref{sql14}) denotes the total detection efficiency,  which accounts for the photon collection efficiency, the splitting into separate detection paths, optical losses, and the detector's quantum efficiency~\cite{gieseler12x}. It also includes the efficiency of translating a displacement in $y$ direction  into a phase change between the excitation field and the scattered field~\cite{gittes98}. The factor $2/5$ stems from the dipolar radiation pattern and is also present in the power spectral density of Eq.~(\ref{coh05}). This factor is identical for the particle's displacement in $z$ direction but gets modified to $1/5$ for the $x$ direction (direction of polarization).\\

To bring the particle close to its quantum ground state we have to minimize the expression in brackets in Eq.~(\ref{sql14}). Without feedback ($S_y^{^{\rm fb}}=0$) the minimum is reached when imprecision noise equals backaction noise, which occurs for
\begin{eqnarray}
P_{\rm scatt}^{(min)}  \;=\;  \sqrt{\frac{5}{8\:\!\eta_c}} \frac{\Omega_0}{\omega_0} \, m c^2\:\!\gamma  \;  \; .
 \label{sql16}
\end{eqnarray}
The total displacement noise corresponding to $P_{\rm scatt}^{(min)}$ turns out to be
\begin{eqnarray}
{\rm Min}\left[S_y(\Omega_0)\right]  \;=\;  S_y^{^{\rm zp}}\!(\Omega_0)/
\sqrt{2 \,/\, 5\:\!\eta_c} \; , 
   \label{sql19}
\end{eqnarray}
which states that the zero-point can be reached if the total detection efficiency is $\eta_c\ge 2/5$. \\

In our experiments we measure the single-sided power spectral density $\tilde{S}_y(f)$, which is related to the mean-square displacement and the double-sided power spectral density $S_y(\Omega)$  as
\begin{equation}
 \left\langle y^2\right\rangle \;=\; \int_{0}^{\infty} \tilde{S}_y(f)\;df\;=\; \int_{-\infty}^{\infty} S_y(\Omega)\;d\Omega\; .
\label{calib03}
\end{equation}
Consequently, $\tilde{S}_{y}(f)=  4\pi \, S_{y}(2\pi f)$. The Lorentzian curve labeled with $n=63$ in Fig. 2 of the main text features an imprecision background of $\tilde{S}_y^{^{\rm imp}} = 1.50\,$pm$^2$/Hz and exhibits a peak amplitude (against the background) of $\tilde{S}_y^{^{\rm peak}}\!(f_0)= 14.53\,$pm$^2$/Hz. Using the value of $\tilde{S}_y^{^{\rm imp}}$  in expression~(\ref{sql14}) together with our experimental parameters  ($\lambda=1064\,$nm, $P_{\rm scatt}=3.53\,\mu$W, $m= 1.14\times 10^{-18}\,$kg, $\Omega_0= 2\pi \times150.030\,$kHz, $\gamma = 2\pi\times 269.9\,$Hz) we determine a total detection efficiency of $\eta_c=0.0005$. \\

\begin{figure}[!t]
\begin{center}
\includegraphics[width=0.7\textwidth]{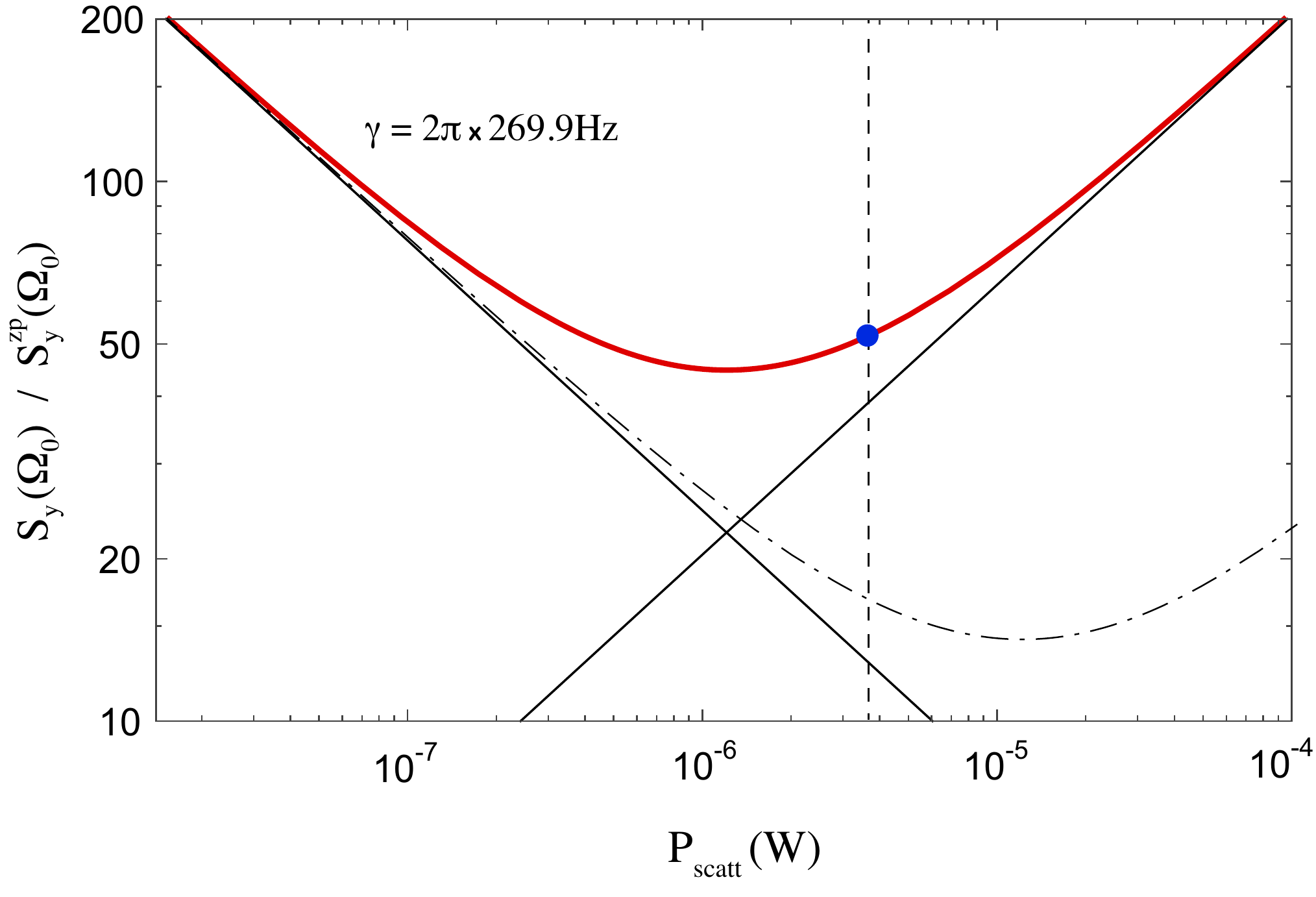}
\end{center}
\quad\vspace{-2.5em}
\caption{Total power spectral density $S_y$ evaluated on resonance ($\Omega=\Omega_0$) as a function of scattered power $P_{\rm scatt}$ for the case of $S_y^{^{\rm fb}}=0$ and for $\eta_c=0.0005$. $S_y$ is normalized by the zero-point spectral density $S_y^{^{\rm zp}}$.  The linewidth of $\gamma = 2\pi\times 269.9\,$Hz is set by the feedback gain. The blue dot corresponds to our experimental conditions ($P_{\rm scatt}=3.53\,\mu$W). The two diagonal lines indicate the contributions of measurement noise and backaction noise, respectively. The dash-dotted curve shows the spectral density $S_y$ for a  linewidth and detection efficiency that are both increased by a factor of ten. 
\label{figsupp01}}
\end{figure}
In Fig.~\ref{figsupp01} we plot the total spectral density $S_y(\Omega_0)$ as a function of scattered power $P_{\rm scatt}$ for the case of $S_y^{^{\rm fb}}=0$ and for $\eta_c=0.0005$. The blue dot on the solid curve indicates our experimental situation, that is, $P_{\rm scatt}=3.53\,\mu$W. Strategies for improved center-of-mass cooling are optimizing the detection efficiency $\eta_c$ and using higher feedback gains (higher $\gamma$). As an example, the dash-dotted curve in Fig.~\ref{figsupp01} shows the case for a linewidth and detection efficiency that are both increased by a factor of ten. \\

According to Fig.~\ref{figsupp01}, the backaction noise at $P_{\rm scatt}=3.53\,\mu$W amounts to $\tilde{S}_y^{^{\rm back}} = 4\pi\times 38.22\times S_y^{^{\rm zp}}= 4.43\,$pm$^2$/Hz, which is a factor of $3.3$ lower than the the experimentally measured peak amplitude of  $\tilde{S}_y^{^{\rm peak}}= 14.53\,$pm$^2$/Hz. We hence conclude that the noise introduced by the feedback loop amounts to $S_y^{^{\rm fb}}=\tilde{S}_y^{^{\rm peak}}-\tilde{S}_y^{^{\rm back}}=10.10\,$pm$^2$/Hz. Thus,  reaching lower center-of-mass temperatures requires that feedback noise be reduced in our parametric cooling scheme.

\newpage


\quad

\end{widetext}

\end{document}